\documentstyle[mprocl]{article}

\def\be{\begin{equation}}
\def\ee{\end{equation}}
\def\bea{\begin{eqnarray}}
\def\eea{\end{eqnarray}}

\begin{document}

\title{Stochastic gravitational waves backgrounds: a probe for inflationary 
and non-inflationary cosmology\footnote{ Based on talks presented
in ``Current issues in string cosmology'' (Paris, June `99), 
``Connecting fundamental physics and cosmology''(Cambridge, August `99) and 
COSMO 99 (Trieste, September `99). To  appear in the Proc. of COSMO 99} }

\author{Massimo Giovannini}

\address{Institute for Theoretical Physics, Lausanne University,\\ 
BSP-Dorigny, CH-1015, Lausanne, Switzerland\\
E-mail: Massimo.Giovannini@ipt.unil.ch}   

\maketitle
\abstracts{Physical scenarios, leading to highly energetic 
stochastic gravitational waves backgrounds (for frequencies 
ranging from the $\mu$Hz up to the GHz) are examined.
In some cases the typical amplitude of the 
logarithmic energy spectrum can be even eight orders of magnitude larger 
than the ordinary inflationary prediction. 
Scaling violations in the frequency dependence of the energy density 
of the produced gravitons are discussed.}

\section{Inflationary graviton spectra and their scaling properties}

The fraction of critical energy 
density $\rho_c$ stored in relic gravitons at 
the present (conformal) time $\eta_0$ 
per each logarithmic interval of the physical frequency $f$
\be
\Omega_{{\rm GW}}(f,\eta_0)\,=\,\frac{1}{\rho_{c}}\,
\frac{{\rm d} \rho_{{\rm GW}}}{{\rm d} \ln{f}}\,=\, 
\overline{\Omega}(\eta_0)\,\omega(f,\eta_0)\,
\label{Omegath}
\ee
is the quantity we will be mostly interested in.
The frequency dependence in $\Omega_{\rm GW}(f,\eta_0)$ is a 
specific feature of the mechanism responsible 
for the production of the gravitons and, in
 a given interval of the present frequency, the slope of the 
logarithmic energy spectrum can be defined as \cite{gio1}
\be
\alpha = \frac{ d\, \ln{\omega(f,\eta_0)}}{d\, \ln{f}}.
\label{sl}
\ee
If, in a given logarithmic interval of frequency, $\alpha <0 $ 
the spectrum is {\em red} since 
its dominant energetical content is stored in the infra-red. 
If, on the other hand $0<\alpha \leq 1$ the spectrum is 
{\em blue}, namely a mildly increasing logarithmic energy 
density. Finally if $ \alpha > 1$ we will talk about {\em violet} 
spectrum whose dominant energetical content is stored in the 
ultra-violet. The case $\alpha =0$ corresponds to the 
case of scale-invariant (Harrison-Zeldovich) logarithmic 
energy spectrum.

Every variation of the background geometry produces 
graviton pairs which are stochastically distributed \cite{gr}. 
The amplitude 
of the detectable signal depends, however, upon the 
specific model of curvature evolution. In ordinary 
inflationary models the amount of gravitons produced 
by a variation of the geometry is notoriously 
quite small \cite{gio2}. This feature can be traced back to the 
fact that $\Omega_{\rm GW}(f,\eta_0)$ is either a decreasing or 
(at most) a flat function of the present frequency. 
Suppose, for simplicity, that the ordinary inflationary phase is 
suddenly followed by a radiation dominated phase turning, after 
some time, into a matter dominated stage of expansion \cite{gio2}. The 
logarithmic energy spectrum will have, as a function of the present 
frequency, two main branches : an infra-red branch (roughly 
ranging between $10^{-18}$ Hz and $10^{-16}$ Hz) and a flat (or 
possibly decreasing) branch between $10^{-16}$ and $100$ MHz. 

The flat (or, more precisely, slightly decreasing) branch of 
the spectrum is mainly due to those modes leaving the 
horizon during the inflationary phase and re-entering 
during the radiation dominated epoch. The infra-red branch of 
the spectrum is  produced by modes leaving the 
horizon during the inflationary phase and re-entering during the 
matter dominated epoch.

Starting from infra-red we have that the COBE observations of the 
first thirty multipole moments of the temperature 
fluctuations in the microwave sky imply that the GW contribution to the 
Sachs-Wolfe integral cannot exceed the amount of anisotropy 
directly detected. This implies that for frequencies $f_0$ 
approximately comparable with $H_0$ and 20 $H_0$ (where 
$H_0$ is the present value of the Hubble constant including 
its indetermination $h_0$) $ h_0^2 \, \Omega_{\rm GW} ( f_0 , \eta_0) 
< 7 \times 10^{-9} $.  
Moving towards the ultra-violet, the very small size of the 
fractional timing error in the arrivals of the 
millisecond pulsar's pulses requires that 
$\Omega_{\rm GW}(f_{P}, \eta_0) < 10^{-8}$ for a typical frequency 
roughly comparable with the inverse of the observation time 
during which the pulses have been monitored , i.e. $f_{P} \sim 10$ nHz.

Finally, if we believe the simplest (homogeneous and isotropic) big-bang
nucleosynthesis (BBN) scenario we have to require that the total fraction of 
critical energy density stored in relic gravitons at the BBN time 
does not exceed the energy density stored in relativistic matter at the 
same epoch. Defining $\Omega_{\gamma}(\eta_0)$ as the fraction of critical 
energy density presently stored in radiation we have that 
the BBN  consistency requirement demands 
\be 
h^2_0\,\int^{f_{\rm max}}_{f_{\rm ns}}\,\Omega_{\rm GW}(f,\eta_0)\;{\rm d}\ln{f}\,<\,0.2\,
h_0^2\,\Omega_{\gamma}(\eta_0)\,\simeq\,5\,\times\,10^{-6},
\label{bbn}
\ee
where  $f_{\rm ns}\,\simeq\,0.1$ nHz is the present value 
of the frequency corresponding to the horizon at the nucleosynthesis 
time; $f_{\rm max}$ stands for  the maximal frequency of the spectrum 
and it depends upon the specific theoretical model (
in the case of ordinary inflationary models $f_{\rm max} = 100$ MHz).
The constraint expressed in Eq. (\ref{bbn}) is 
{\em global} in the sense that it bounds the {\em integral} of the 
logarithmic energy spectrum. The constraints coming from 
pulsar's timing errors and from the integrated Sachs-Wolfe effect
are instead {\em local} in the sense that they 
bound the value of the logarithmic energy spectrum in a specific interval of 
frequencies. 

In the case of stochastic GW backgrounds of 
 inflationary origin, owing to the red nature
of the logarithmic energy spectrum, the most significant constraints 
are the ones present in the soft region of the spectrum, 
more specifically, the 
ones connected with the Sachs-Wolfe effect. 
Taking into account the specific frequency behavior in the 
infra-red branch of the spectrum and assuming perfect scale invariance 
we have that $h_0^2\,\,\,\Omega_{\rm GW} ( f,\eta_0) < 10^{-15}$ for 
frequencies $f> 10^{-16}$ Hz. 
We have to conclude that 
the inflationary spectra are invisible by pairs of interferometric detectors 
operating in a 
window ranging approximately between few Hz and $10$ kHz. 

In order to illustrate more quantitatively this point we remind the 
expression of the signal-to-noise ratio (SNR)  in the 
context of optimal processing  required for the detection of stochastic backgrounds 
\cite{int}. By 
assuming that the intrinsic 
 noises of the detectors are stationary, gaussian, 
uncorrelated, much larger in amplitude than the gravitational strain, and 
statistically independent on the strain itself, one has: 
\begin{equation}
{\rm SNR}^2 \,=\,\frac{3 H_0^2}{2 \sqrt{2}\,\pi^2}\,F\,\sqrt{T}\,
\left\{\,\int_0^{\infty}\,{\rm d} f\,\frac{\gamma^2 (f)\,\Omega^2_{{\rm GW}}(f)}
{f^6\,S_n^{\,(1)} (f)\,S_n^{\,(2)} (f)}\,\right\}^{1/2}\; ,
\label{2}
\end{equation}
($F$ depends upon 
the geometry of the two detectors and in the case of the correlation between 
two interferometers $F=2/5$; $T$ is the observation time). 
In Eq. (\ref{2}), $S_n^{\,(k)} (f)$ is the (one-sided) noise power 
spectrum (NPS) of the $k$-th 
$(k = 1,2)$ detector. The NPS contains the important informations concerning the 
noise sources (in broad terms seismic, thermal and shot noises)
 while $\gamma(f)$ is the overlap reduction function 
which is determined by the relative locations and orientations 
of the two detectors. Without going through the technical details 
\cite{gio3}
from the expression of the SNR we want to notice that the 
achievable sensitivity of a pair of wide band interferometers crucially 
depends upon the spectral slope of the theoretical energy spectrum in the 
operating window of the detectors. So, a flat spectrum will lead 
to an experimental sensitivity which might not be similar to the 
sensitivity achievable in the case of a blue or violet spectra 
\cite{gio3,gio4}. In the case of an exactly scale invariant spectrum
the correlation of the two (coaligned) LIGO detectors with 
central corner stations in Livingston (Lousiana) and in Hanford 
(Washington) will have a sensitivity to a flat spectrum 
which is $h_0^2\,\,\, \Omega_{\rm GW}(100~{\rm Hz}) \simeq 6.5 \times 10^{-11} $ 
after one year of observation and with signal-to-noise 
ratio equal to one \cite{gio3}. 
This implies that ordinary inflationary spectra 
are (and will be) invisible by wide band detectors since the 
inflationary prediction, in the most favorable case (i.e. scale invariant spectra), 
undershoots the experimental sensitivity by more than four orders of magnitude.

\section{Scaling violations in graviton spectra}

In order to have  a large detectable signal between $1$ Hz and $10$ kHz we have to 
look for models exhibiting scaling violations for frequencies larger than 
the mHz. The scaling violations should go in the direction 
of blue ($0< \alpha \leq 1$) or violet  ($\alpha > 1$) 
logarithmic energy spectra. Only in this case we shall have the hope that 
the signal will be large enough in the window 
of wide band detectors. Notice that the growth of the spectra should 
not necessarily be monotonic: we might have a blue or violet spectrum for 
a limited interval of frequencies with a spike or a hump.

\subsection{Quintessential inflationary models}

Suppose now, as a toy example, that the ordinary inflationary phase is not immediately 
followed by a radiation dominated phase but by a quite long phase 
expanding slower than radiation \cite{gio1}. This speculation  is theoretically 
plausible since we ignore what was the thermodynamical history of the Universe 
prior to BBN. If the Universe expanded slower than radiation the equation of state 
of the effective sources driving the geometry had to be, for some time, 
stiffer than radiation. This means that the effective speed of sound $c_s$ 
had to lie in the range $1/\sqrt{3} < c_{s} \leq 1$.
Then the resulting logarithmic energy spectrum,  for the 
modes leaving the horizon during the inflationary phase and 
re-entering during the stiff phase, is tilted towards large 
frequencies with typical (blue) slope given by \cite{gio1}
\begin{equation}
\alpha = \frac{ 6 c_s^2 - 2 }{3 c_s^2 +1}\,,\,\,\,\,\, 0< \alpha \leq 1.
\end{equation}
A situation very similar to the one we just described occurs in 
quintessential inflationary models \cite{vil1}. In this case the 
tilt is maximal (i.e., $\alpha=1 $) and a more precise calculation shows 
the appearance of logarithmic corrections in the logarithmic 
energy spectrum which becomes \cite{gio1,vil1,gio4} 
$\omega(f) \propto f \ln^2{f}$.
The maximal frequency $f_{\rm max}(\eta_0)$ is of the order of $100$ GHz
(to be compared with the $100$ MHz of ordinary inflationary models)
and it corresponds to 
the typical frequency of a  spike in the GW background. In quintessential 
inflationary models the relic graviton background will then have 
the usual infra-red and flat branches supplemented, at high 
frequencies (larger than the mHz and smaller than the GHz) by a true 
hard branch \cite{gio4} whose peak can be, in terms of $h_0^2 \,\,\, 
\Omega_{\rm GW}$, 
 of the order of $10^{-6}$, compatible 
with the BBN bound and  roughly eight orders of magnitude larger than the 
signal provided by ordinary inflationary models. 

An interesting aspect of this class of models is that the maximal signal 
occurs in a frequency region between the MHz and the GHz. 
Microwave cavities can be used as GW detectors precisely in the 
mentioned frequency range \cite{cav1}. There were published results 
reporting the construction  of this type of detectors \cite{cav2} and the 
possibility of further improvements in the sensitivity received 
recently attention \cite{cav3}. Our signal is certainly a candidate 
for this type of devices.

\subsection{String cosmological models} 

In string cosmological models \cite{ven} of pre-big-bang type 
$h_0^2 \,\,\,\Omega_{\rm GW}$ can 
be as large as $10^{-7}$--$10^{-6}$ for frequencies ranging between $1$ Hz and 
$100$ GHz \cite{gio2,gas1}. 
In these types of models the logarithmic energy spectrum can be 
either blue or violet depending upon the given mode of the spectrum. If the mode
under consideration 
left the horizon during the dilaton-dominated epoch the typical 
slope will be violet (i.e. $ \alpha \sim 3 $ up to logarithmic corrections).
If the given mode left the horizon during the stringy phase the slope can 
be also blue with typical spectral slope $\alpha \sim 6 - 2 ( \ln{g_1/g_s}/\ln{z_s})$
where $g_1$ and $g_s$ are the values of the dilaton coupling at the 
end of the stringy phase and at the end of the dilaton dominated phase; $z_s$ 
parametrizes the duration of the stringy phase. This 
behaviour is representative of the minimal string cosmological 
scenarios. However, in the non-minimal case the spectra can also be non 
monotonic \cite{gas1}. Recently the sensitivity of a pair of VIRGO detectors
to string cosmological gravitons was specifically analyzed \cite{gio10} with the 
conclusion that a VIRGO pair, in its upgraded stage, will certainly be able to probe 
wide regions of the parameter space of these models. If we  maximize the 
overlap between the two detectors \cite{gio10} or 
if we would  reduce (selectively) the pendulum and pendulum's internal modes
contribution to the thermal noise of the instruments \cite{gio11}, the 
visible region (after one year of observation and with SNR equal to one)
of the parameter space will get even larger. Unfortunately, as in the 
case of the advanced LIGO detectors, also in the case of the advanced VIRGO 
detector the sensitivity to a flat spectrum will be irrelevant for 
ordinary inflationary models. Finally, it is worth mentioning 
that blue and violet logarithmic energy spectra can also arise 
in the context of other models like dimensional decoupling \cite{gio5} 
and early violations of the dominant energy condition \cite{gio6} in 
Einsteinian theories of gravity.  

\section{Relic gravitons from local processes inside the horizon}

GW can be produced not only because of the adiabatic variation of the 
background geometry, but also because there are physical processes 
occurring inside the horizon producing large amounts of gravitational 
radiation. Typical examples of such a statement are topological defects models, 
strongly first order phase transitions (where the bubble collisions can produce 
spikes in the GW background for frequencies roughly 
comparable with the inverse of horizon/bubble size).
For instance, if the EWPT would be strongly first order we would have spikes in the 
graviton background for frequencies between $10$ $\mu$Hz and $0.1$ mHz.
In the following we want to discuss a further mechanism connected with 
the existence of hypermagnetic fields in the symmetric phase of the electroweak 
theory \cite{shap}. 

\subsection{ Magnetic and Hypermagnetic Knots}

Since  a generic magnetic field configuration at finite conductivity 
leads to an 
energy-momentum tensor which is anisotropic and which has non-vanishing 
transverse and traceless component (TT), 
if magnetic fields are present inside the horizon at some epoch they can 
radiate 
gravitationally.
The TT components of the energy momentum tensor act as a source term for the 
TT fluctuations of the geometry which are associated with 
gravitational waves. A non-trivial example of this effect is 
provided by magnetic knot configurations \cite{gio7} which are 
transverse (magnetic) field configurations with 
a topologically non-trivial structure in the flux lines. These 
configurations can also be generated by direct projection of a 
pure $SU(2)$ field onto a fixed (electromagnetic) direction in isospace
\cite{jak}. 
In magnetohydrodynamics (MHD) magnetic knots configurations 
are stable and conserved by plasma evolution provided 
the conductivity is sufficiently large.
The degree of knottedness of the configuration 
is measured by the magnetic helicity. 
Assuming a specific configuration \cite{gio7} 
the frequency of the hump in the GW spectrum 
could range between $10^{-14}$ Hz and $10^{-12}$ Hz. 

For sufficiently high temperatures and for sufficiently large values of the various 
fermionic charges the $SU(2)_{L}\otimes U(1)_{Y}$ symmetry is restored and, thence, 
non-screened vector modes will now correspond to the hypercharge group. 
Topologically non-trivial configurations of the hypermagnetic field 
($\vec{{\cal H}}_{Y}$)
 can be related to the 
baryon asymmetry of the Universe (BAU) \cite{shap,laine} and they can also radiate 
gravitationally \cite{gio8,rub}. The evolution 
equations of the hypercharge field at finite conductivity 
imply that the largest modes which can survive in the plasma 
are the ones associated with the hypermagnetic conductivity 
frequency which is roughly eight orders of magnitude smaller 
than the temperature at the time of the electroweak 
phase transition which I take to occur  
around $100$ GeV. The logarithmic energy spectra of the 
produced gravitons can be different depending upon the 
specific form of the configuration. However, we can estimate
\be
h_0^2 \Omega_{\rm GW}(f, \eta_0) \simeq 10^{-6} \,\, \delta^4 ,  
\ee
where $\delta = |\vec{{\cal H}}_{Y}|/T^2_{\rm ew}$ and $T_{\rm ew}$ roughly 
corresponds to $100$ GeV. Notice that $\delta \sim 1 $ does not violate the 
closure density bound since it sould be divided by $N_{\rm eff}$ (i.e. the 
effective number of spin degrees of freedom at $T_{\rm ew}$) which is 
already included in the numerical prefactor of our estimate.
The frequency 
$f$ lies in the range between $10$ $\mu$Hz and the few  kHz. 
The lower frequency  corresponds to the frequency of the horizon at the 
electroweak epoch, i.e. 
\be
f_{\rm ew}(\eta_0) \sim 0.201 \,\,
\biggl(\frac{T_{\rm ew}}{1~{\rm GeV}}\biggr)~\biggl(\frac{N_{eff}}{100}\biggr)^{1/6}
 \mu{\rm Hz}.
\ee
The higher frequency roughly corresponds to the hypermagnetic conductivity 
frequency, namely $f_{\sigma}(\eta_0) \sim 10^{8} ~ f_{\rm ew}(\eta_0)$.
The presence of a classical hypermagnetic background in the 
symmetric phase of the electroweak theory produces interesting 
non linear effects in the phase diagram of the 
electroweak phase transition \cite{laine}. If we then suppose 
\cite{laine} that $\delta > 0.3$ we can get 
$h_0^2\,\,\, \Omega_{\rm GW}$  as large as $10^{-7}$.
This signal satisfies the above mentioned 
phenomenological bounds on the graviton backgrounds of 
primordial origin \cite{gio8}.

\section{Final remarks}
In spite of the fact that ordinary inflationary models provide 
a rather minute relic graviton background, we showed that 
various physical situations can provide a much larger signal. 
A pair of correlated and coaligned 
VIRGO detectors (for ${\rm Hz} < f < 10 \,\,\, 10 {\rm kHz}$)
 and microwave cavities (for ${\rm MHz} < f < {\rm GHz}$ )
can offer exciting detectability prospects.

\section*{Acknowledgments}

I would like to thank D. Babusci, L. Ford,  M. Gasperini, R. Jackiw, 
M. Laine, M. Shaposhnikov, P. Steinhardt, P. Tinyakov, N. Turok, 
G. Veneziano and A. Vilenkin for extremely useful 
discussions. I wish also to thank 
E. Coccia, G. Pizzella and E. Picasso for informative discussions.
Finally, I wish to thank the organizers of 
 the  COSMO 99 meeting, and in particular G. Senjanovic, for 
their kind invitation.

\end{document}